\documentclass{amsart}
\usepackage{graphicx}

\theoremstyle{definition}

\theoremstyle{remark}

\numberwithin{equation}{section}

\allowdisplaybreaks
\usepackage[utf8]{inputenc}
\theoremstyle{plain}
\usepackage{amsmath}
\usepackage{amsfonts}
\usepackage{amssymb}
\usepackage{graphicx}
\usepackage{lscape}
\usepackage{subcaption}
\usepackage{booktabs}
\usepackage{multicol}
\usepackage[all,cmtip]{xy}
\usepackage{algorithmic}
\usepackage[linesnumbered,ruled]{algorithm2e}
\usepackage{vaucanson-g}
\usepackage[T1]{fontenc}
\usepackage{float}
\usepackage{tikz}
\usepackage{makecell}
\usepackage{babel,blindtext}
\usetikzlibrary{arrows}
\newtheorem{Theorem}{Theorem}[section]      

\newtheorem{Conjecture}[Theorem]{Conjecture}
\newtheorem{Proposition}[Theorem]{Proposition}
\newtheorem{Remark}[Theorem]{Remark}
\newtheorem{Proof}[Theorem]{Proof}
\newtheorem{Example}[Theorem]{Example}

\newtheorem{Lemma}[Theorem]{Lemma}

\theoremstyle{remark}

\numberwithin{equation}{section}

\begin{document}

\title{A Continued Fraction-Hyperbola based Attack on RSA cryptosystem}

\author{Gilda Rech Bansimba}
\address{Department of Mathematics and Computer Science\\ Universit\'e Marien Ngouabi \\ Facult\'e des Sciences et Techniques \\ BP: 69, Brazzaville, Congo}
\curraddr{Department of Mathematics and Computer Science\\ Universit\'e Marien Ngouabi \\ Facult\'e des Sciences et Techniques \\ BP: 69, Brazzaville, Congo}
\email{bansimbagilda@gmail.com}

\author{Regis F. Babindamana}
\address{Department of Mathematics and Computer Science\\ Universit\'e Marien Ngouabi \\ Facult\'e des Sciences et Techniques \\ BP: 69, Brazzaville, Congo}
\email{regis.babindamana@umng.cg}
%
\author{Basile Guy R. Bossoto}
\address{Department of Mathematics Universit\'e Marien Ngouabi \\ Facult\'e des Sciences et Techniques \\ BP: 69, Brazzaville, Congo}
\email{basile.bossoto@umng.cg}



\begin{abstract} \ \\ \rm{
In this paper we present new arithmetical and algebraic results following the work of Babindamana and al. on hyperbolas and describe in the new results an approach to attacking a RSA-type modulus based on continued fractions, independent and not bounded by the size of the private key $d$ nor the public exponent $e$ compared to Wiener's attack. When successful, this attack is bounded by $\displaystyle\mathcal{O}\left( b\log{\alpha_{j4}}\log{(\alpha_{i3}+\alpha_{j3})}\right)$ with $b=10^{y}$, $\alpha_{i3}+\alpha_{j3}$ a non trivial factor of $n$ and $\alpha_{j4}$ such that $(n+1)/(n-1)=\alpha_{i4}/\alpha_{j4}$. The primary goal of this attack is to find a point $\displaystyle X_{\alpha}=\left(-\alpha_{3}, \ \alpha_{3}+1 \right) \in \mathbb{Z}^{2}_{\star}$ that satisfies $\displaystyle\left\langle X_{\alpha_{3}}, \ P_{3} \right\rangle =0$ from a convergent of $\displaystyle\frac{\alpha_{i4}}{\alpha_{j4}}+\delta$, with $P_{3}\in \mathcal{B}_{n}(x, y)_{\mid_{x\geq 4n}}$. We finally present some experimental examples. We believe these results constitute a new direction in RSA Cryptanalysis using continued fractions independently of parameters $e$ and $d$.
}
\end{abstract}
\maketitle
\section*{Introduction}
In \cite{Gilda}, Babindamana and al. presented arithmetical results on hyperbola and applied them to lattice points with the Fermat factorization Method. In particular, they used results from the hyperbola parametrization $\mathcal{B}_{n}(x, y)_{\mid_{\mathbb{Z}}}$ to prove that for the Fermat factorization equation if $(N-6)\equiv 0 \mod 4$, the Fermat's method fails. Otherwise, in terms of cardinality, it has respectively $4$, $8$, $2(\alpha+1)$, $(1-\delta_{2p_{i}})2^{n+1}$ and $2\prod_{i=1}^{n}(\alpha_{i}+1)$ lattice points respectively if $N$ is an odd prime, $N=N_{a}\times N_{b}$ with $N_{a}$ and $N_{b}$ odd primes, $N=N_{a}^{\alpha}$ with $N_{a}$ prime, $N=\prod_{i=1}^{n}p_{i}$ with $p_{i}$ distinct primes and $N=\prod_{i=1}^{n} N_{i}^{\alpha_{i}}$ with $N_{i}$ odd primes. \\ In this paper, we present new arithmetical results for the hyperbola parametrization considered in \cite{Gilda}, in particular we prove that 
given $p$, $a$, $\alpha_{i}$, $\alpha_{j}\in \mathbb{Z}_{>0}$ with $\gcd(p+1, p-1) \neq 1$, if $\gcd(\alpha_{i}, \alpha_{j})=1$ and $\displaystyle\frac{p+1}{p-1}=\frac{\alpha_{i}}{\alpha_{j}}$ then $\alpha_{i}+\alpha_{j}=p$ and 
\begin{eqnarray*}
	\alpha_{i}^{a}-\alpha_{j}^{a} =
	\left\lbrace
	\begin{array}{ll}
	1 \ \ if \ a=1 \\
	p \ \ if \ a=2 \\
	p\delta \ \ if \ 2\mid a
	\end{array} \right.
\end{eqnarray*}
 Additionally, among other results, given a RSA modulus $n$, we prove that for all $P_{i}\in \mathcal{B}_{n}(x, y)_{\mid_{x\geq 4n}}$, there exists $\alpha_{i}\in \mathbb{Z}_{\geq 0}$ such that\\ $P_{i}=\displaystyle\left( 4n\frac{(\alpha_{i}+1)^{2}} {2\alpha_{i}+1}, \ 4n(\alpha_{i}+1)\cdot \frac{\alpha_{i}}{2\alpha_{i}+1}\right)$, with $\alpha_{0}=0$, $\alpha_{4}=n-1$, $\alpha_{i}$ an integer sequence and $\displaystyle\left\lbrace P_{2}, \ P_{3}, \ P_{4} \right\rbrace $ the algebraic subset of $\mathcal{B}_{n}(x, y)_{\mid_{x\geq 4n}}$. Finally, we present an attack on those results using continued fractions whose goal is to find $X_{\alpha}$ such that $\displaystyle\left\langle X_{\alpha_{3}}, \ P_{3} \right\rangle =0$. 
When successful, this attack has a complexity of\\ $\displaystyle\mathcal{O}\left( b\log{\alpha_{j4}}\log{(\alpha_{i3}+\alpha_{j3})}\right)$ with $b=10^{y}$, $\alpha_{i3}+\alpha_{j3}$ a non trivial factor of $n$ and $\alpha_{j4}$ such that $(n+1)/(n-1)=\alpha_{i4}/\alpha_{j4}$.
\section{New algebraic results on $\mathcal{B}_{N}(x, y)_{\mid_{x\geq 4N}}$}
In \cite{Rech}, the authors presented some arithemtical results on hyperbola and in \cite{Gilda}, Babindamana and al. presented other arithmetical and algebraic results on the hyperbola $\mathcal{B}_{n}(x, y)_{\mid_{\mathbb{Z}}}$ and their application to the lattice points of the Fermat factorization Method.\\
Recall that $\mathcal{B}_{n}(x, y)_{\mid_{\mathbb{Z}}}=\displaystyle \lbrace \left(x, y\right)\in \mathbb{Z}\times \mathbb{Z} \ / \ y^{2}=x^{2}-4nx \displaystyle \rbrace$ and $\mathcal{B}_{n}(x, y)_{\mid_{x\geq 4n}}= \lbrace (x, y)\in \mathbb{Z}_{\geq 4n}\times \mathbb{Z}_{\geq 0} \ / \ y^{2}=x^{2}-4nx\rbrace$.\\
In this section, we present new algebraic results on $\mathcal{B}_{N}(x, y)_{\mid_{x\geq 4N}}$.
\begin{Theorem} \ \\ \rm{
Let $p$, $a$, $\alpha_{i}$, $\alpha_{j}\in \mathbb{Z}_{>0}$ with $\gcd(p+1, p-1) \neq 1$. If $\gcd(\alpha_{i}, \alpha_{j})=1$ and $\displaystyle\frac{p+1}{p-1}=\frac{\alpha_{i}}{\alpha_{j}}$ then $\alpha_{i}+\alpha_{j}=p$ and 
\begin{eqnarray*}
	\alpha_{i}^{a}-\alpha_{j}^{a} =
	\left\lbrace
	\begin{array}{ll}
	1 \ \ if \ a=1 \\
	p \ \ if \ a=2 \\
	p\delta \ \ if \ 2\mid a
	\end{array} \right.
\end{eqnarray*}
}
\end{Theorem}
\begin{Proof} \ \\ \rm{
\begin{itemize}
\item Consider $p\in \mathbb{Z}\setminus \lbrace0, \ 1 \rbrace$ with $\gcd(p+1, p-1) \neq 1$, then the rational fraction $\displaystyle\frac{p+1}{p-1}$ is reducible. Now Assume $\displaystyle\frac{p+1}{p-1}=\frac{\alpha_{i}}{\alpha_{j}}$ its irreducible expression, where $\gcd(\alpha_{i}, \alpha_{j})=1$. Rearranging the terms in this expression, we have $p\alpha_{j}+\alpha_{j}=p\alpha_{i}-\alpha_{i}$, then $-p(\alpha_{i}-\alpha_{j})=-(\alpha_{i}+\alpha_{j})$ hence $p=\displaystyle\frac{\alpha_{i}+\alpha_{j}}{\alpha_{i}-\alpha_{j}}$. Now since by assumption, $p$ is an integer, then $(\alpha_{i}-\alpha_{j})\mid (\alpha_{i}+\alpha_{j})$ which is true only if $\alpha_{i}-\alpha_{j}$=1 since $\gcd(p+1, p-1) \neq 1$.\\
Assume $\alpha_{i}-\alpha_{j}\neq 1$, $p+1=\displaystyle\frac{2\alpha_{i}}{\alpha_{i}-\alpha_{j}}$ and $p-1=\displaystyle\frac{2\alpha_{j}}{\alpha_{i}-\alpha_{j}}$. In this case, $\gcd(p+1, \ p-1)=\gcd(\displaystyle\frac{2\alpha_{i}}{\alpha_{i}-\alpha_{j}}, \ \displaystyle\frac{2\alpha_{j}}{\alpha_{i}-\alpha_{j}})$. Since the greatest common divisor is defined only in an Euclidean domain, then this would imply that $(\alpha_{i}-\alpha_{j})\mid 2\alpha_{i}$ and $(\alpha_{i}-\alpha_{j})\mid 2\alpha_{j}$, which is absurd since by hypothesis, $\gcd(\alpha_{i}, \alpha_{j})=1$ and by applying the lemma of Gauss, if $(\alpha_{i}-\alpha_{j})\mid 2\alpha_{i}$ then $(\alpha_{i}-\alpha_{j})\nmid 2\alpha_{j}$ or conversely if $(\alpha_{i}-\alpha_{j})\mid 2\alpha_{j}$ then $(\alpha_{i}-\alpha_{j})\nmid 2\alpha_{i}$, hence in this case, the $gcd$ is not defined.\\
Therefore, this holds only if $\alpha_{i}-\alpha_{j}=1$.
\item If $a=1$, since $p=\displaystyle\frac{\alpha_{i}+\alpha_{j}}{\alpha_{i}-\alpha_{j}}$ then $\alpha_{i}-\alpha_{j}=1$
\item If $a=2$, $\alpha_{i}^{2}-\alpha_{j}^{2}=(\alpha_{i}-\alpha_{j})(\alpha_{i}+\alpha_{j})=1\times p=p$.
\item Considering the Newton's binomial formula $\displaystyle(\alpha_{i}+\alpha_{j})^{n}=\sum_{k=0}^{n}\binom{n}{k}\alpha_{i}^{k}\alpha_{j}^{n-k}$, assume $2\mid a$, then there exists $\beta \in \mathbb{Z}$ such that $a=2\beta$. 
In this case
\begin{align*}
\alpha_{i}^{a}-\alpha_{j}^{a}&=\alpha_{i}^{2\beta}-\alpha_{j}^{2\beta}=(\alpha_{i}^{2})^{\beta}-(\alpha_{j}^{2})^{\beta}\\
 &=(\alpha_{i}^{2}-\alpha_{j}^{2})(\alpha_{i}^{2(\beta-1)}+\alpha_{i}^{2(\beta-2)}\alpha_{j}^{2}+ \ \cdots \ + \alpha_{i}^{2}\alpha_{j}^{2(\beta-2)}+\alpha_{j}^{2(\beta-1)})\\
 &=p(\alpha_{i}-\alpha_{j})\left(\sum_{k=0}^{\beta-1}\alpha_{i}^{2(\beta-1-k)}\alpha_{j}^{2k} \right)\\
 &=p\left(\sum_{k=0}^{\beta-1}\alpha_{i}^{2(\beta-1-k)}\alpha_{j}^{2k} \right)
\end{align*}
Setting $\displaystyle\delta=\left(\sum_{k=0}^{\beta-1}\alpha_{i}^{2(\beta-1-k)}\alpha_{j}^{2k} \right)$, we have $\alpha_{i}^{a}-\alpha_{j}^{a}=p\delta$
\end{itemize}
}
\end{Proof}
\begin{Theorem} \ \\ \ \rm{
If $\delta$ is a perfect square of a prime, then $Card\displaystyle\left( \mathcal{B}_{\delta}(x, y)_{\mid_{x\geq 4\delta}}\right)=3$.
}
\end{Theorem}
\begin{Proof} \ \\ \rm{
$\delta$ is a perfect square with prime $t$ such that $\delta=t^{2}$. In this case, $Div(\delta)=\lbrace 1, t, \delta \rbrace$, in other words, $\delta$ has 3 divisors.\\
From theorem 2 in \cite{Gilda}, using injective homomorphisms, we have 
the trivial injective morphism $f_{1\longrightarrow \delta}$, the trivial automorphism $f_{\delta\longrightarrow \delta}$, and the morphism $f_{t\longrightarrow \delta}$. with $\mathcal{B}_{1}(x, y)_{\mid_{x\geq 4}}\lbrace (4, 0)\rbrace$, $\mathcal{B}_{t}(x, y)_{\mid_{x\geq 4t}}\lbrace (4, 0), \ ((t+1)^{2}, t^{2}-1)\rbrace$. $\mathcal{B}_{1}(x, y)_{\mid_{x\geq 4}}=\lbrace (4, 0)\rbrace$ and $\mathcal{B}_{t}(x, y)_{\mid_{x\geq 4t}}=\lbrace (4t, 0), \ ((t+1)^2, t^2-1)\rbrace$ given by, for all $i\in Div(\delta)$
\begin{eqnarray*}
f_{i\rightarrow \delta}: \ \ \mathcal{B}_{i}(x, y)_{\mid_{x\geq 4i}} \ \ \ \longrightarrow \ \ \ \ \mathcal{B}_{\delta}(x, y)_{\mid_{x\geq 4\delta}} \\ 
(x, y) \ \ \longmapsto \ \displaystyle\left( x\frac{\delta}{i}, y\frac{\delta}{i}) \right)
\end{eqnarray*}
with $f_{i\rightarrow \delta}((4, 0))=(4\delta, \ 0) \in \mathcal{B}_{\delta}(x, y)_{\mid_{x\geq 4\delta}}$, $f_{t\rightarrow \delta}((4, 0))=(4\delta, \ 0) \in \mathcal{B}_{\delta}(x, y)_{\mid_{x\geq 4\delta}}$, $f_{t\rightarrow \delta}(((t+1)^{2}, t^2-1))=(t(t+1)^{2}, \ t(t^2-1)) \in \mathcal{B}_{\delta}(x, y)_{\mid_{x\geq 4\delta}}$ and $((\delta+1)^2, \delta^2-1) \in \mathcal{B}_{\delta}(x, y)_{\mid_{x\geq 4\delta}}$.\\ Hence $\in \mathcal{B}_{\delta}(x, y)_{\mid_{x\geq 4\delta}}$ has exactly 3 points. 

}
\end{Proof}

\begin{Theorem} \ \label{principal} \rm{
\begin{enumerate}
\item Let $n$ be a RSA modulus, $\alpha_{i}$ an integer sequence and $\displaystyle\left\lbrace P_{0}, \ P_{2}, \ P_{3}, \ P_{4} \right\rbrace $ the algebraic subset of $\mathcal{B}_{n}(x, y)_{\mid_{x\geq 4n}}$.\\
There exists $X_{\alpha_{i}}=(-\alpha_{i}, \ \alpha_{i}+1)\in \mathbb{Z}^{2}$, such that $\displaystyle\left\langle X_{\alpha_{i}}, \ P_{i} \right\rangle_{i=\overline{0, 2\cdots 4}} =0$ and $2\alpha_{i}+1 \ \in \lbrace 1,  p, q, n \rbrace$.
\item For $P_{2}$ and $P_{3}$ with $\displaystyle\left\langle X_{\alpha_{2}}, \ P_{2} \right\rangle =0$ and $\displaystyle\left\langle X_{\alpha_{3}}, \ P_{3} \right\rangle =0$, then $\displaystyle\frac{x_{P_{1}}}{y_{P_{1}}}=\frac{\alpha_{3}+\alpha_{2}+1}{\alpha_{3}-\alpha_{2}}$
\item $p$, $q$ respectively $n$ and $\varphi(n)$ are given by $2\alpha_{2}+1$, $2\alpha_{3}+1$ respectively $2\alpha_{4}+1$ and $4\alpha_{2}\alpha_{3}$.

\item \begin{eqnarray*}
	\forall \ P_{i} \in \displaystyle\left\lbrace P_{0}, P_{1}, P_{2}, P_{3}, P_{4} \right\rbrace \text{, } \ \  P_{i} =
	\left\lbrace
	\begin{array}{ll}
	x_{P_{i}}\displaystyle\left(1, \ \frac{\alpha_{3}-\alpha_{2}}{\alpha_{3}+\alpha_{2}+1} \right)  \ \ if \ i=1 \\
	x_{P_{i}}\displaystyle\left(1, \ \frac{\alpha_{i}}{\alpha_{i}+1} \right)  \ \ else 
	\end{array} \right.
\end{eqnarray*}
\item \begin{align*}
P_0&=\left(16\alpha_{2}\alpha_{3}+8\alpha_{2}+8\alpha_{3}+4, \ 0\right)\\
P_1&=\left(4 \, \alpha_{2}^{2} + 8 \, \alpha_{2} \alpha_{3} + 4 \, \alpha_{3}^{2} + 8 \, \alpha_{2} + 8 \, \alpha_{3} + 4, \ \ 4 \, \alpha_{2}^{2} - 4 \, \alpha_{3}^{2} + 4 \, \alpha_{2} - 4 \, \alpha_{3}\right)\\
P_2&=\left(8 \, \alpha_{2} \alpha_{3}^{2} + 16 \, \alpha_{2} \alpha_{3} + 4 \, \alpha_{3}^{2} + 8 \, \alpha_{2} + 8 \, \alpha_{3} + 4, \ \ 8 \, \alpha_{2} \alpha_{3}^{2} + 8 \, \alpha_{2} \alpha_{3} + 4 \, \alpha_{3}^{2} + 4 \, \alpha_{3}\right)\\
P_3&=\left(8 \, \alpha_{2}^{2} \alpha_{3} + 4 \, \alpha_{2}^{2} + 16 \, \alpha_{2} \alpha_{3} + 8 \, \alpha_{2} + 8 \, \alpha_{3} + 4, \ \ 8 \, \alpha_{2}^{2} \alpha_{3} + 4 \, \alpha_{2}^{2} + 8 \, \alpha_{2} \alpha_{3} + 4 \, \alpha_{2}\right)\\
P_4&=\left(4 \, \alpha_{4}^{2} + 8 \, \alpha_{4} + 4, \ \ 4 \, \alpha_{4}^{2} + 4 \, \alpha_{4}\right)
\end{align*}
\item the ${\alpha_{i}}_{\vert_{i=\overline{2,\cdots 4}}}$s as defined in $(1)$ satisfy the following system:
\begin{eqnarray*}
	\left\lbrace
	\begin{array}{ll}
	4\alpha_{2}\alpha_{3}+2(\alpha_{2}+\alpha_{3})+1=n \ \ \ \ \ \ (a) \\ \\
	4(16\alpha_{2}^{3}\alpha_{3}^{3}+32\alpha_{2}^{3}\alpha_{3}^{2}+32\alpha_{2}^{2}\alpha_{3}^{3}+20\alpha_{2}^{3}\alpha_{3}+68\alpha_{2}^{2}\alpha_{3}^{2}+20\alpha_{2}\alpha_{3}^{3}-2\alpha_{2}^{2}\alpha_{3}n-2\alpha_{2}\alpha_{3}^{2}n+\\46\alpha_{2}^{2}\alpha_{3}+46\alpha_{2}\alpha_{3}^{2}+4\alpha_{3}^{3}-\alpha_{2}^{2}n-8\alpha_{2}\alpha_{3}n-\alpha_{3}^{2}n+10\alpha_{2}^{2}+34\alpha_{2}\alpha_{3}+\\10\alpha_{3}^{2}-4\alpha_{2}n-4\alpha_{3}n+n^2+8\alpha_{2}+8\alpha_{3}-2n+2)=n(n+1)^2 \ \ \ \ \ \ (b)\\ \\
4(8\alpha_{2}^{2}\alpha_{3}^{2}+12\alpha_{2}^{2}\alpha_{3}+12\alpha_{2}\alpha_{3}^{2}+4\alpha_{2}^{2}+18\alpha_{2}\alpha_{3}+4\alpha_{3}^{2}-\\ \alpha_{2}n-\alpha_{3}n+6\alpha_{2}+6\alpha_{3}-n+2)(2\alpha_{2}\alpha_{3}+\alpha_{2}+\alpha_{3})=n(n^2-1) \ \ \ \ \ \ (c)
\end{array} \right.
\end{eqnarray*}

or equivalently, setting $P=\alpha_{2}\alpha_{3}$ and $S=\alpha_{2}+\alpha_{3}$, $P$ and $S$ satisfy the following system:
\begin{eqnarray*}
	\left\lbrace
	\begin{array}{ll}
	4P+2S+1=n \ \ \ \ \ \ (a) \\ \\
	16P^{3}+4S^{3}+(32S+28)P^{2}+(20P+10-n)S^{2}+(34-2n)PS+\\(14-6n)P+(8-4n)S=\displaystyle\frac{n(n+1)^{2}}{4}-n^{2}+2n-2 \ \ \ \ \ \ (b) \\ \\
	\left[8P^{2}+4S^{2}+12PS+10P+(6-n)S-n+2 \right](2P+S)=\displaystyle\frac{n(n^2-1)}{4}   \ \ \ \ \ \ (c)

\end{array} \right.
\end{eqnarray*}
\item For all $i\in \lbrace 0, 2, 3, 4  \rbrace$, $\displaystyle\frac{\alpha_{ai}}{\alpha_{bi}}<\frac{\alpha_{aj}}{\alpha_{bj}}$ $\forall \ j>i$.
\item More generally, there exists $\alpha_{i}\in \mathbb{Z}_{\geq 0}$ such that\\ $P_{i}=\displaystyle\left( 4n\frac{(\alpha_{i}+1)^{2}} {2\alpha_{i}+1}, \ 4n(\alpha_{i}+1)\cdot \frac{\alpha_{i}}{2\alpha_{i}+1}\right) \in \mathcal{B}_{n}(x, y)_{\mid_{x\geq 4n}}$

\item Considering $\mathcal{B}_{n}(x, y)_{\mid_{x\geq 4n}}$ for a RSA modulus $n$, the general term of the sequence $\alpha_{n}$ is $\alpha_{i}=\displaystyle\frac{\tau_{i} -1}{2} \ \ \forall \ \tau_{i} \in Div(n)$, with $\alpha_{0}=0$ and $\alpha_{4}=\frac{n-1}{2}$.
\end{enumerate}
}
\end{Theorem}

\begin{Proof} \ \\ \ \rm{
\begin{enumerate}
\item To prove this, let's consider $X_{\alpha_{i}}=(-\alpha_{i}, \ \alpha_{i}+1)\in \mathbb{Z}^{2}$ and $P_{i}=\left(x_{P_{i}}, \ y_{P_{i}} \right)$ with $i\in\left\lbrace 0, 2, 3, 4 \right\rbrace $. The scalar product $\displaystyle\left\langle X_{\alpha_{i}}, \ P_{i} \right\rangle_{i=\overline{0, 2\cdots 4}}=0$ is equivalent to $-\alpha_{i}x_{P_{i}}+(\alpha_{i}+1)y_{P_{i}}$=0, finally $\alpha_{i}=\displaystyle\frac{y_{P_{i}}}{x_{P_{i}}-y_{P_{i}}}$.\\ Here without loss of generality, for all $P_{i} \in \displaystyle\left\lbrace P_{0}, \ P_{2}, \ P_{3}, \ P_{4} \right\rbrace$, from $\mathcal{B}_{n}(x, y)_{\mid_{x\geq 4n}}$ structure, $x_{P_{i}} \neq y_{P_{i}}$, hence $\alpha_{i}$ is well defined and\\  $X_{\alpha_{i}}=\displaystyle \left(-\frac{y_{P_{i}}}{x_{P_{i}}-y_{P_{i}}}, \ \frac{x_{P_{i}}}{x_{P_{i}}-y_{P_{i}}} \right) $.
By the same, 
\begin{align*} 
2\alpha_{i}+1&=\displaystyle\frac{2y_{P_{i}}}{x_{P_{i}}-y_{P_{i}}}+1 \ \ \ \ \ \ \ \ \ \ \ \ \ \ \ \ \ \ \ \ \ \ \ \ \ \ \ \ \ \ \ \ \ \ \ \ \ \ \ \ \ \ \ \ \ \ \ \ \ \ \ \ \ \ \ \ \ \ \ \ \ \ \ \ \  \\
&=\displaystyle\frac{x_{P_{i}}+y_{P_{i}}}{x_{P_{i}}-y_{P_{i}}}=\frac{(x_{P_{i}}+y_{P_{i}})^{2}}{x_{P_{i}}^{2}-y_{P_{i}}^{2}}=\displaystyle\frac{x_{P_{i}}^{2}+y_{P_{i}}^{2}+2x_{P_{i}}y_{P_{i}}}{x_{P_{i}}^{2}-y_{P_{i}}^{2}}\\
&=\displaystyle\frac{x_{P_{i}}^{2}+x_{P_{i}}^{2}+2x_{P_{i}}y_{P_{i}} -4nx_{P_{i}}}{x_{P_{i}}^{2}-x_{P_{i}}^{2}+4nx_{P_{i}}}\text{ since } y_{P_{i}}^{2}=x_{P_{i}}^{2}-4nx_{P_{i}}^{2}\\
&= \displaystyle\frac{2x_{P_{i}}^{2}+2x_{P_{i}}y_{P_{i}} -4nx_{P_{i}}}{4nx_{P_{i}}}\\
&=\displaystyle\frac{x_{P_{i}}}{2n}+\displaystyle\frac{y_{P_{i}}}{2n}-1 \text{, since for all $P_{i} \in \displaystyle\left\lbrace P_{0}, \ P_{2}, \ P_{3}, \ P_{4} \right\rbrace$, }\\
&P_{i} \in \left\lbrace \left( 4n, \ 0\right), \ \left( p(q+1)^{2}, \ p(q^{2}-1)\right), \ \left( q(p+1)^{2}, \ q(p^{2}-1)\right)\right\rbrace\\&\cup \left\lbrace\left( (n+1)^{2}, \ n^{2}-1\right) \right\rbrace\\
\text{for each $P_{i}$, we have:}\\
\displaystyle\frac{x_{P_{i}}}{2n}+\displaystyle\frac{y_{P_{i}}}{2n}-1&=\displaystyle\frac{4n+0}{2n}-1=2-1=1 \  \in\left\lbrace 1, p, q, n\right\rbrace \text{ or } \\
\displaystyle\frac{x_{P_{i}}}{2n}+\displaystyle\frac{y_{P_{i}}}{2n}-1&=\displaystyle\frac{p(q+1)^{2}+p(q^{2}-1)}{2n}-1=\displaystyle\frac{nq+2n+p+qn-p-2n}{2n}=q \\
&\ \in\left\lbrace 1, p, q, n\right\rbrace \text{ or } \\
&=\displaystyle\frac{q(p+1)^{2}+q(p^{2}-1)}{2n}-1=\displaystyle\frac{np+2n+q+pn-q-2n}{2n}=p \\
& \ \in\left\lbrace 1, p, q, n\right\rbrace \text{ or } \\
&=\displaystyle\frac{(n+1)^{2}+n^{2}-1}{2n}-1=\displaystyle\frac{n^2+2n+1+n^2-1}{2n}=\displaystyle\frac{2n^2}{2n}=n \\
& \text{Hence } 2\alpha_{i}+1 \ \in \lbrace 1, p, q, n \rbrace \text{ for all } i\in\lbrace 0, 2, 3, 4 \rbrace.
\end{align*}
\item Here, under the same conditions as above, consider $2\alpha_{2}+1=p$ and $2\alpha_{3}+1=q$. From $\mathcal{B}_{n}(x, y)_{\mid_{x\geq 4n}}$ structure, 
\begin{align*}
\displaystyle\frac{x_{P_{1}}}{y_{P_{1}}}&=\frac{(p+q)^{2}}{p^{2}-q^{2}}=\frac{p+q}{p-q} \ \ \ \ \ \ \ \ \ \ \ \ \ \ \ \ \ \ \ \ \ \ \ \ \ \ \ \ \ \ \ \ \ \ \ \ \ \ \ \ \ \ \ \ \ \ \ \ \ \ \ \ \ \ \ \ \ \ \ \ \ \ \ \ \ \ \ \ \ \ \ \ \ \ \ \ \ \ \ \ \ \ \ \ \ \\
&=\frac{2\alpha_{2}+1+2\alpha_{3}+1}{2\alpha_{2}+1-2\alpha_{3}-1}\\
&=\frac{\alpha_{2}+\alpha_{3}+1}{\alpha_{2}-\alpha_{3}}.
\end{align*}
\item This results from \textit{$(2)$} for $p$ and $q$, and for the Euler totient function $\varphi(n)=(p-1)(q-1)=\alpha_{2}\alpha_{3}$.
\item If $i=1$,  from \textit{(2)}, $\displaystyle\frac{x_{P_{1}}}{y_{P_{1}}}=\frac{\alpha_{2}+\alpha_{3}+1}{\alpha_{2}-\alpha_{3}}$, then $y_{P_{1}}=\displaystyle\frac{\alpha_{3}-\alpha_{2}}{\alpha_{3}+\alpha_{2}+1}$, hence $\left(x_{P_{1}}, \ \displaystyle\frac{\alpha_{3}-\alpha_{2}}{\alpha_{3}+\alpha_{2}+1}x_{P_{1}} \right)$.\\
If not $i=1$, considering the scalar product relation from (\textit{1}), $\displaystyle\left\langle X_{\alpha_{i}}, \ P_{i} \right\rangle_{i=\overline{2\cdots 4}} =0$, we have $\displaystyle\frac{x_{P_{i}}}{y_{P_{i}}}=\frac{\alpha_{i}+1}{\alpha_{i}}$ then $y_{P_{1}}=\displaystyle\frac{\alpha_{i}}{\alpha_{i}+1}$. In this case, for all $i \in \lbrace 2, 3, 4 \rbrace$, $P_{i}=\displaystyle\left(x_{P_{i}}, \ y_{P_{i}} \right)=\displaystyle\left(x_{P_{i}}, \ \frac{\alpha_{i}}{\alpha_{i}+1}x_{P_{i}}\right)$. This holds for $i=0$. Hence 
\begin{eqnarray*}
	\forall \ P_{i} \in \displaystyle\left\lbrace P_{0}, P_{1}, P_{2}, P_{3}, P_{4} \right\rbrace \text{, } \ \  P_{i} =
	\left\lbrace
	\begin{array}{ll}
	x_{P_{i}}\displaystyle\left(1, \ \frac{\alpha_{3}-\alpha_{2}}{\alpha_{3}+\alpha_{2}+1} \right)  \ \ if \ i=1 \\
	x_{P_{i}}\displaystyle\left(1, \ \frac{\alpha_{i}}{\alpha_{i}+1} \right)  \ \ else 
	\end{array} \right.
\end{eqnarray*}
\item From $(1)$, $2\alpha_{2}+1=p$, $2\alpha_{3}+1=q$ and $2\alpha_{4}+1=n$, then substituting these values in $\mathcal{B}_{n}(x, y)_{\mid_{x\geq 4n}}=$\\ 
\begin{scriptsize}
$\left\lbrace \left( 4n, \ 0\right), \ \left( (p+q)^2, \ p^2-q^2\right), \ \left( p(q+1)^{2}, \ p(q^{2}-1)\right), \ \left( q(p+1)^{2}, \ q(p^{2}-1)\right), \ \left( (n+1)^{2}, \ n^{2}-1\right) \right\rbrace$
\end{scriptsize}
For $P_{0}=(4n, \ 0)$, we get $P_0=\left(16\alpha_{2}\alpha_{3}+8\alpha_{2}+8\alpha_{3}+4, \ 0\right)$\\
For $P_{1}$, $P_{2}$, $P_{3}$ and $P_{4}$, after expansion, we get\\ $P_1=\left(4 \, \alpha_{2}^{2} + 8 \, \alpha_{2} \alpha_{3} + 4 \, \alpha_{3}^{2} + 8 \, \alpha_{2} + 8 \, \alpha_{3} + 4, \ \ 4 \, \alpha_{2}^{2} - 4 \, \alpha_{3}^{2} + 4 \, \alpha_{2} - 4 \, \alpha_{3}\right)$; $P_2=\left(8 \, \alpha_{2} \alpha_{3}^{2} + 16 \, \alpha_{2} \alpha_{3} + 4 \, \alpha_{3}^{2} + 8 \, \alpha_{2} + 8 \, \alpha_{3} + 4, \ \ 8 \, \alpha_{2} \alpha_{3}^{2} + 8 \, \alpha_{2} \alpha_{3} + 4 \, \alpha_{3}^{2} + 4 \, \alpha_{3}\right)$; $P_3=\left(8 \, \alpha_{2}^{2} \alpha_{3} + 4 \, \alpha_{2}^{2} + 16 \, \alpha_{2} \alpha_{3} + 8 \, \alpha_{2} + 8 \, \alpha_{3} + 4, \ \ 8 \, \alpha_{2}^{2} \alpha_{3} + 4 \, \alpha_{2}^{2} + 8 \, \alpha_{2} \alpha_{3} + 4 \, \alpha_{2}\right)$ and $P_4=\left(4 \, \alpha_{4}^{2} + 8 \, \alpha_{4} + 4, \ \ 4 \, \alpha_{4}^{2} + 4 \, \alpha_{4}\right)$
\item For $(a)$, since $n=(2\alpha_{2}+1)(2\alpha_{3}+1)=4\alpha_{2}\alpha_{3}+2(\alpha_{2}+\alpha_{3})+1$.\\
Now for $(b)$ and $(c)$, we know from \cite{Gilda}, that $P_{2}+P_{3}=P_{4}$.\\ Since $\displaystyle\left(x_{P_{2}+P_{3}}, \ y_{P_{2}+P_{3}}\right)=\left(x_{P_{4}}, \ y_{P_{4}}\right)$. We set $\displaystyle\left( (b), \ (c) \right)=\displaystyle\left(x_{P_{2}+P_{3}}, \ y_{P_{2}+P_{3}}\right)$.\\
Recall the additive group law on $\mathcal{B}_{n}(x, y)_{\mid_{\mathbb{Z}}}$ for points $P_{2}$ and $P_{3}$ is defined by
\begin{eqnarray*}
\left\lbrace
\begin{array}{ll}
x_{P_{2}+P_{3}}=\displaystyle\frac{1}{2n}\left[(x_{P_{2}}-2n)(x_{P_{3}}-2n)+y_{P_{2}}y_{P_{3}}\right]+2n\\
y_{P_{2}+P_{3}}=\displaystyle\frac{1}{2n}\left[y_{P_{2}}(x_{P_{3}}-2n)+y_{P_{3}}(x_{P_{2}}-2n)\right]
\end{array} \right.
\end{eqnarray*}
Then substituting the expressions of $P_{2}$ and $P_{3}$ from $(3)$ in terms of $\alpha_{i}^{'}$s in the system  
\begin{eqnarray*}
\left\lbrace
\begin{array}{ll}
\displaystyle\frac{1}{2n}\left[(x_{P_{2}}-2n)(x_{P_{3}}-2n)+y_{P_{2}}y_{P_{3}}\right]+2n=x_{P_{4}}=(n+1)^{2}\\
\displaystyle\frac{1}{2n}\left[y_{P_{2}}(x_{P_{3}}-2n)+y_{P_{3}}(x_{P_{2}}-2n)\right]=y_{P_{4}}=n^{2}-1
\end{array} \right.
\end{eqnarray*}
we finally get  
\begin{eqnarray*}
	\left\lbrace
	\begin{array}{ll}
	4(16\alpha_{2}^{3}\alpha_{3}^{3}+32\alpha_{2}^{3}\alpha_{3}^{2}+32\alpha_{2}^{2}\alpha_{3}^{3}+20\alpha_{2}^{3}\alpha_{3}+68\alpha_{2}^{2}\alpha_{3}^{2}+20\alpha_{2}\alpha_{3}^{3}-2\alpha_{2}^{2}\alpha_{3}n-2\alpha_{2}\alpha_{3}^{2}n+\\46\alpha_{2}^{2}\alpha_{3}+46\alpha_{2}\alpha_{3}^{2}+4\alpha_{3}^{3}-\alpha_{2}^{2}n-8\alpha_{2}\alpha_{3}n-\alpha_{3}^{2}n+10\alpha_{2}^{2}+34\alpha_{2}\alpha_{3}+\\10\alpha_{3}^{2}-4\alpha_{2}n-4\alpha_{3}n+n^2+8\alpha_{2}+8\alpha_{3}-2n+2)=n(n+1)^2 \ \ \ \ \ \ (b)\\ \ \ \ \ \ \ \ \ \ \ \ \ \ \ \ \ \ \ \ \ \ \ \ \ \ \ \ \ \ \ \ \ \ \ \ \ \ \ \ \ \ \ \ \ \ \ \ \ \ \ \ \ \ \ \ \ \ \ \ \ \ \ \ \ \ \ \ \ \ \ \ \ \ \ \ \ \ \ \ \ \ \ \ \ \ \ \ \ \ \ \ \ \ \ \ \ \ \ \ \ \ \ \ \ \  \ \  (6.2)\\
4(8\alpha_{2}^{2}\alpha_{3}^{2}+12\alpha_{2}^{2}\alpha_{3}+12\alpha_{2}\alpha_{3}^{2}+4\alpha_{2}^{2}+18\alpha_{2}\alpha_{3}+4\alpha_{3}^{2}-\\ \alpha_{2}n-\alpha_{3}n+6\alpha_{2}+6\alpha_{3}-n+2)(2\alpha_{2}\alpha_{3}+\alpha_{2}+\alpha_{3})=n(n^2-1) \ \ \ \ \ \ (c)
\end{array} \right.
\end{eqnarray*}
setting $P=\alpha_{2}\alpha_{3}$, $S=\alpha_{2}+\alpha_{3}$, and substituting in both the relation $(a)$ and the system $(6.2)$ above, we get:
\begin{eqnarray*}
	\left\lbrace
	\begin{array}{ll}
	4P+2S+1=n \ \ \ \ \ \ (a) \\ \\
	16P^{3}+4S^{3}+(32S+28)P^{2}+(20P+10-n)S^{2}+(34-2n)PS+\\(14-6n)P+(8-4n)S=\displaystyle\frac{n(n+1)^{2}}{4}-n^{2}+2n-2 \ \ \ \ \ \ (b) \\ \\
	\left[8P^{2}+4S^{2}+12PS+10P+(6-n)S-n+2 \right](2P+S)=\displaystyle\frac{n(n^2-1)}{4}   \ \ \ \ \ \ (c)

\end{array} \right.
\end{eqnarray*}
\item Here first recall that $\displaystyle\left( \alpha_{ai}, \ \alpha_{bi}\right)=\left(\alpha_{ai}, \ \alpha_{ai}+1 \right)$ and $\displaystyle\left( \alpha_{aj}, \ \alpha_{bj}\right)=\displaystyle\left(\alpha_{aj}, \ \alpha_{aj}+1 \right)$. If $j>i$, then $\displaystyle\alpha_{j}>\alpha_{i}$ considering $(1)$. Now $\displaystyle\alpha_{i}<\alpha_{j}$ then $\displaystyle\alpha_{i}+\alpha_{i}\alpha_{j}<\alpha_{j}+\alpha_{i}\alpha_{j}$ since $\alpha_{i}>0$,  $\alpha_{j}>0$. We have $\displaystyle\alpha_{i}\left(\alpha_{j}+1\right)<\alpha_{j}\left(\alpha_{i}+1\right)$. Finally $\displaystyle\frac{\alpha_{i}}{\alpha_{i}+1}<\frac{\alpha_{j}}{\alpha_{j}+1}$. Hence $\displaystyle\frac{\alpha_{ai}}{\alpha_{bi}}<\frac{\alpha_{aj}}{\alpha_{bj}}$.
\item  Consider $P_{i}=\displaystyle\left( 4n\frac{(\alpha_{i}+1)^{2}} {2\alpha_{i}+1}, \ 4n(\alpha_{i}+1)\cdot \frac{\alpha_{i}}{2\alpha_{i}+1}\right) \in \mathcal{B}_{n}(x, y)_{\mid_{x\geq 4n}}$,
since from $(1)$, $2\alpha_{i}+1 \ \in \lbrace 1,  p, q, n \rbrace$ then $\alpha_{i} \in \displaystyle\left\lbrace 0, \frac{p-1}{2}, \frac{q-1}{2}, \frac{n-1}{2}\right\rbrace $ then:\\
If $\alpha_{i}=0$, $P_{i}=\displaystyle\left(4n\frac{1}{1}, \ 4n\cdot(1)\cdot 0 \right)=\displaystyle\left(4n, \ 0\right) \in \mathcal{B}_{n}(x, y)_{\mid_{x\geq 4n}}$.\\
If $\alpha_{i}=\displaystyle\frac{p-1}{2}$, $P_{i}=\displaystyle\left(4n\cdot\frac{\left(\frac{p+1}{2}\right)^{2}}{p}, \ 4n\cdot\frac{p+1}{2}\cdot \frac{\frac{p-1}{2}}{p} \right)$\\   \hspace*{2.64cm}$=\displaystyle\left(4n\cdot\frac{(p+1)^{2}}{4p}, \ 4n\cdot\frac{p+1}{2}\cdot\frac{p-1}{2p} \right)$\\
 \hspace*{2.64cm}$=\displaystyle\left(q(p+1)^{2}, \ q(p^{2}-1)\right) \in \mathcal{B}_{n}(x, y)_{\mid_{x\geq 4n}}$.\\
If $\alpha_{i}=\displaystyle\frac{q-1}{2}$, $P_{i}=\displaystyle\left(4n\cdot\frac{\left(\frac{q+1}{2}\right)^{2}}{q}, \ 4n\cdot\frac{q+1}{2}\cdot \frac{\frac{q-1}{2}}{q} \right)$\\  \hspace*{2.64cm}$=\displaystyle\left(4n\cdot\frac{(q+1)^{2}}{4q}, \ 4n\cdot\frac{q+1}{2}\cdot\frac{q-1}{2q} \right)$\\
 \hspace*{2.64cm}$=\displaystyle\left(p(q+1)^{2}, \ p(q^{2}-1)\right) \in \mathcal{B}_{n}(x, y)_{\mid_{x\geq 4n}}$\\
If $\alpha_{i}=\frac{n-1}{2}$, $P_{i}=\displaystyle\left(4n\cdot\frac{\left(\frac{n+1}{2}\right)^{2}}{n}, \ 4n\cdot\frac{n+1}{2}\cdot \frac{\frac{n-1}{2}}{n} \right)$ \\ \hspace*{2.64cm}$=\displaystyle\left(4n\cdot\frac{(n+1)^{2}}{4n}, \ 4n\cdot\frac{n+1}{2}\cdot\frac{n-1}{2n} \right)$\\
 \hspace*{2.64cm}$=\displaystyle\left((n+1)^{2}, \ n^{2}-1\right) \in \mathcal{B}_{n}(x, y)_{\mid_{x\geq 4n}}$.
\item This is a straightforward observation from $(1)$. Since $2\alpha_{i}+1 \in \displaystyle\left\lbrace 1, p, q, n\right\rbrace$, then either $\alpha_{i}=0$, $\displaystyle\alpha_{i}=\frac{p-1}{2}$ or $\displaystyle\alpha_{i}=\frac{q-1}{2}$ or $\displaystyle\alpha_{i}=\frac{n-1}{2}$.
 
\end{enumerate}
}
\end{Proof}

\begin{Conjecture} \ \\ \ \rm{
For all $i\in \left\lbrace 1, 2, 3, 4 \right\rbrace$, $P_{i}\in \displaystyle\left\lbrace P_{1}, P_{2}, \ P_{3}, \ P_{4} \right\rbrace $ the algebraic subset of $\mathcal{B}_{n}(x, y)_{\mid_{x\geq 4n}}$. If there exists $(\alpha_{ii}, \alpha_{ji})\in \mathbb{Z}_{\geq 0}^{2}$ such that $\displaystyle\frac{x_{P_{i}}}{y_{P_{i}}}=\frac{\alpha_{ii}}{\alpha_{ji}}$ with $\gcd(\alpha_{ii}, \ \alpha_{ji})=1$ then $\alpha_{i1}+\alpha_{j1}=\alpha_{i3}+\alpha_{j3}$ and $\alpha_{ii}+\alpha_{ji}\in \ \lbrace p, q, n \rbrace$.
}
\end{Conjecture}
\begin{Example} \ \\ \rm{
Here we consider a sample of $11$ different $n$ in the following table, their $\mathcal{B}_{n}(x, y)_{\mid_{x\geq 4n}}$ structures with the corresponding $\displaystyle\frac{\alpha_{ii}}{\alpha_{ji}}$:\\
\begin{center}

\begin{tiny}
\begin{landscape}
 \vspace*{2.5cm}
 \begin{table}[h!]
  \caption{Table of $\mathcal{B}_{n}(x, y)_{\mid_{x\geq 4n}}$ and $\displaystyle\frac{\alpha_{ii}}{\alpha_{ji}}$ for $11$ different $n$ values}
\begin{tabular}{ c c c c c c }
\hline\hline
\thead{n} & \thead{$\mathcal{B}_{n}(x, y)_{\mid_{x\geq 4n}}$} & \thead{$\displaystyle\frac{\alpha_{i1}}{\alpha_{j1}}$} & \thead{$\displaystyle\frac{\alpha_{i2}}{\alpha_{j2}}$} & \thead{$\displaystyle\frac{\alpha_{i3}}{\alpha_{j3}}$} & \thead{$\displaystyle\frac{\alpha_{i4}}{\alpha_{j4}}$} \\ 
\hline 
15 & $\left\lbrace (60, 0), (64, 16), (80, 40), (108, 72), (256, 224) \right\rbrace $ & $4$ & $2$ & $\frac{3}{2}$ & $\frac{8}{7}$ \\ 
35 & $\left\lbrace(140, 0), (144, 24), (252, 168), (320, 240), (1296, 1224) \right\rbrace $ & $6$ & $\frac{3}{2}$ & $\frac{4}{3}$ & $\frac{18}{17}$ \\ 
77 & $\left\lbrace(308, 0), (324, 72), (704, 528), (1008, 840), (6084, 5928) \right\rbrace $ & $\frac{9}{2}$ & $\frac{4}{3}$ & $\frac{6}{5}$ & $\frac{39}{38}$ \\ 
143 & $\left\lbrace(572, 0), (576, 48), (1872, 1560), (2156, 1848), (20736, 20448) \right\rbrace $ & $12$ & $\frac{6}{5}$ & $\frac{7}{6}$ & $\frac{72}{71}$ \\ 
391 & $\left\lbrace (1564, 0), (1600, 240), (7452, 6624), (9792, 8976), (153664, 152880)\right\rbrace $ & $\frac{20}{3}$ & $\frac{9}{8}$ & $\frac{12}{11}$ & $\frac{196}{195}$ \\ 
713 & $\left\lbrace (2852, 0), (2916, 432), (17856, 16368), (23552, 22080), (509796, 508368)\right\rbrace $ & $\frac{27}{4}$ & $\frac{12}{11}$ & $\frac{16}{15}$ & $\frac{357}{356}$ \\ 
14893 & \makecell{$(59572, 0), (111556, 76152), (819396, 789048), (4214772, 4184880),$ \\ $ (221831236, 221801448)$} & $\frac{167}{114}$ & $\frac{27}{26}$ & $\frac{141}{140}$ & $\frac{7447}{7446}$ \\ 
44923 & \makecell{$(179692, 0), (190096, 44472), (7592256, 7501872), (12174300, 12084120),$ \\ $ (2018165776, 2018075928)$} & $\frac{218}{51}$ & $\frac{84}{83}$ & $\frac{135}{134}$ & $\frac{22462}{22461}$ \\ 

58307 & \makecell{$(233228, 0), (242064, 46248), (11720000, 11602800),$ \\ $ (17200764, 17083752), (3399822864, 3399706248) $} & $\frac{246}{47}$ & $\frac{100}{99}$ & $\frac{147}{146}$ & $\frac{29154}{29153}$ \\
439007603 & \makecell{$(1756030412, 0), (2065157136, 798996408), (6116692964112, 6115814885880),$ \\ $ (13835324622476, 13834446579408), (192727676369820816, 192727675491805608) $} & $\frac{22722}{8791}$ & $\frac{6966}{6965}$ & $\frac{15757}{15756}$ & $\frac{219503802}{219503801}$ \\ 
\makecell{12100028890\\760359639} & \makecell{$(48400115563041438556, 0), (49139988961662726400, 6029707342921211040),$\\  $ (37206579369977836429336472656, 37206579345777778639945602360),$ \\ $ (47614527370423585388056617100, 47614527346223527600386064680),$\\$ (14641069915723537932$ \\ $2158145982140929600, 1464106991572353792$ \\ $97958088200620210320)$} & $\frac{3504996040}{430079469}$ & $\frac{1537458286}{1537458285}$ & $\frac{1967537755}{1967537754}$ & $\frac{6050014445380179820}{6050014445380179819}$ \\
\hline 
\end{tabular} 
 \end{table}
\end{landscape}
\end{tiny}
\end{center}
}
\end{Example}
\begin{figure}[!h]
\centering 
\minipage{0.49\textwidth}
\centerline{\fbox{\includegraphics[width=6cm, height=4.5cm]{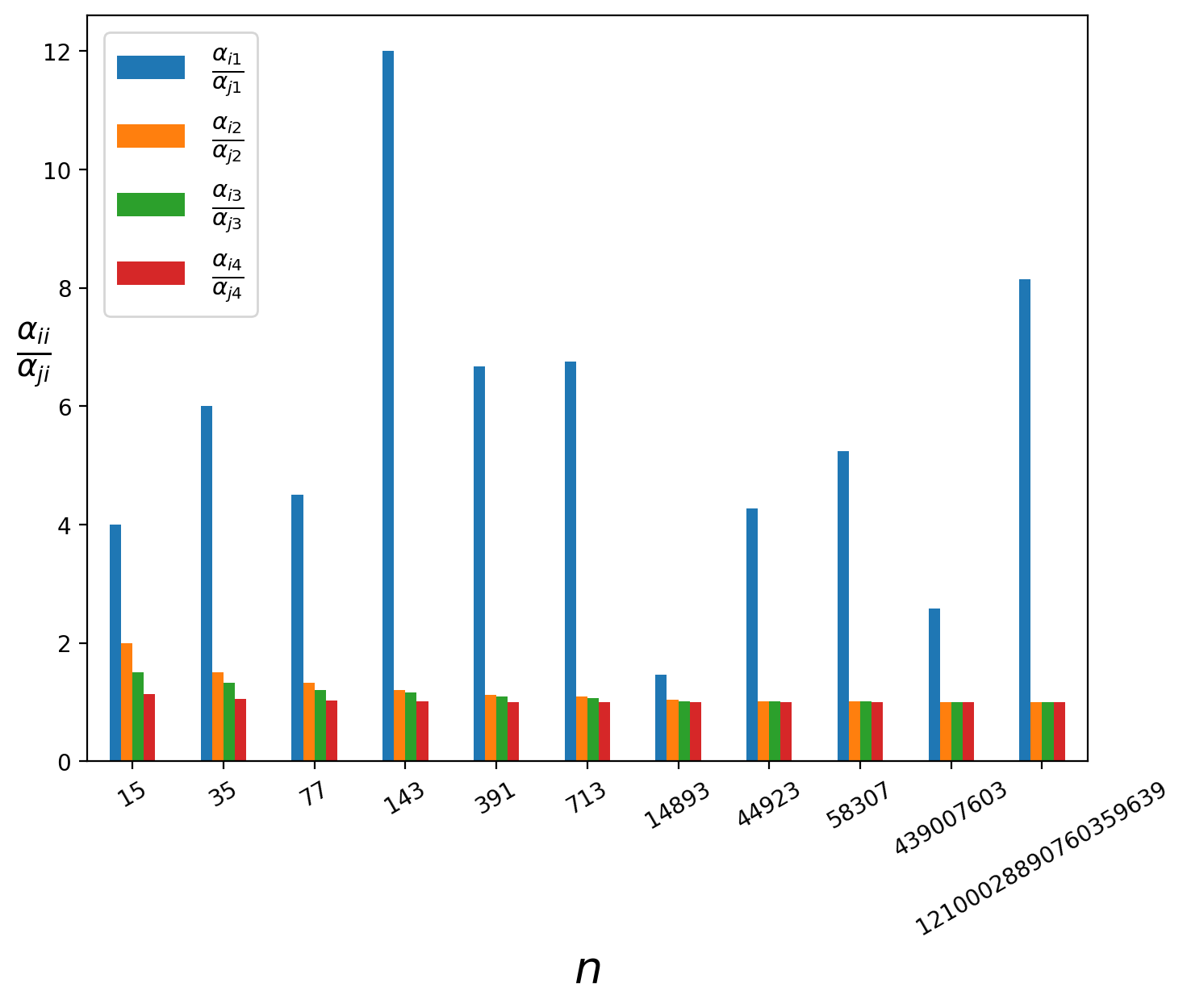}}}
\endminipage\hfill
\minipage{0.51\textwidth}
\centerline{\fbox{\includegraphics[width=6cm, height=4.5cm]{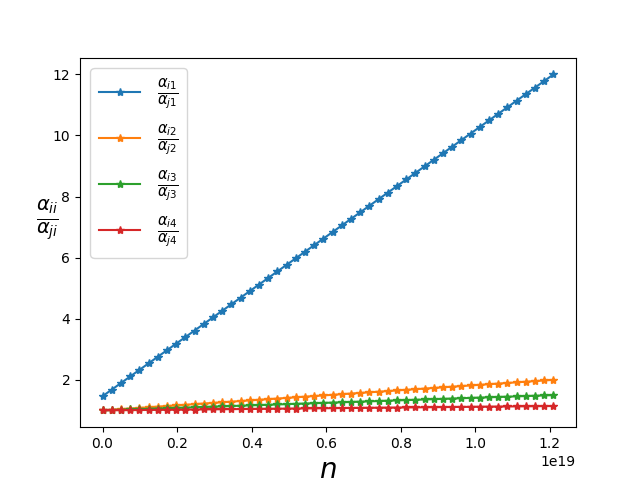}}}
\endminipage\hfill
\caption{$\displaystyle\frac{\alpha_{ii}}{\alpha_{ji}}$ from a sample of $10$ $\mathcal{B}_{n}(x, y)_{\mid_{x\geq 4n}}$ }
\label{bandwidth} 
\end{figure}
From these plots, an observation is that the maximum value of $\displaystyle\frac{\alpha_{ii}}{\alpha_{ji}}$ on $\mathcal{B}_{n}(x, y)_{\mid_{x\geq 4n}}$ is obtained with $i=1$: the point $P_{1}$, followed by the point with $i=2$: the point $P_{2}$ then followed by the $P_{3}$ and finally by the point $P_{4}$. The sequence $\displaystyle\frac{\alpha_{ii}}{\alpha_{ji}}$ is a decreasing sequence.\\
Another notice for this sample is that for $ i\in \lbrace 2, 3, 4\rbrace$, the sequence $\frac{\alpha_{i}}{\alpha_{j}}$ seems to converge as $n$ goes to infinity, and $\displaystyle\frac{\alpha_{i2}}{\alpha_{j2}} \ \simeq \ \displaystyle\frac{\alpha_{i3}}{\alpha_{j3}} \ \simeq \ \displaystyle\frac{\alpha_{i4}}{\alpha_{j4}}$ as $n$ grows.\\

\section{The Attack}
In this section, we present the attack. 
This is a generic attack and only depends on the modulus $n$ and does not depend on the size of the public parameter $e$ or the private exponent $d$ in the cryptosystem setting. This is a new approach compared to Wiener's  attack.\\
Prior to the presentation, let's recall some preliminaries on the RSA cryptosystem and on some non-generic attack related work done so far.
\subsection{Recall on RSA Cryptosystem} \ \rm{
We denote $\mathcal{P}$, $\mathcal{C}$, $\mathcal{K}$, $\mathcal{E}_{k}$, $\mathcal{D}_{k}$ respectively the Plaintext space, the Ciphertext space, the Key space, the Encryption algorithm and the Decryption algorithm. \\
Here $\mathcal{P}=\mathcal{C}=\mathbb{Z}/n\mathbb{Z}$, and\\ $\mathcal{K}=\displaystyle\left\lbrace  (p, q, n, e, d)\in \mathbb{Z}_{+}^{5},\ p, q \text{ primes}\ / \ n=pq, \ ed\equiv 1 \ \mod \varphi(n) \right\rbrace $, $p$, $q$ are chosen randomly, $1< e < \varphi(n) \ / \ \gcd(e, \varphi(n))=1$ and $d$ such that $ed \equiv 1 \mod \varphi(n)$. Public parameters are $(n, e)$, secret parameters are $(p, q, d)$. The following are the encryption and decryption algorithms.
\begin{align*}
\mathcal{E}_{k} \ \ : \ \ \mathcal{P} \ &\longrightarrow \ \mathcal{C} \ \ \ \ \ \ \ \ \ \ \ \ \ \ \ \ \ \ \ \ \ \ \ \ \ \ ; \ \ \ \ \ \ \ \ \ \ \ \ \mathcal{D}_{k} \ \ : \ \ \mathcal{C} \longrightarrow \ \mathcal{P} \\ 
m \ &\longmapsto  \ \mathcal{E}_{k}\displaystyle\left(m \right)=m^{e} \mod n  \ \ \ \ \ \ \ \ \ \ \ \ \ \ \ \ \ \ \ \ \;
\ \ \  c \ \longmapsto  \ \mathcal{D}_{k}\displaystyle\left(c \right)=c^{d} \mod n \ \ \ \ \ \ \ \ \ \ \ \ \ \ \ \ \ \ \ \ \ \ \ \ \ \ \ \ \ \ \ \ \ \ \
\end{align*}
$ \forall \ m \in \mathcal{P}, \ \mathcal{E}_{k}(m)=m^{e} \mod n=c$ and $\mathcal{D}_{k}(c)=c^{d} \mod n=m$. With $\mathcal{D}_{k}\displaystyle\left(\mathcal{E}_{k}\right)=\mathcal{E}_{k}\displaystyle\left(\mathcal{D}_{k}\right)=id$.
}
\subsection{Related work on non-generic attack}
Most of the non-generic attacks are based on the development of continued fractions \cite{Wiener} and the application of Coppersmith's \cite{coppersmith} theorem.
From the work of Wiener in \cite{Wiener}, it is shown that the secret key $d$ can be recovered in polynomial time for a bound of $d<\displaystyle\frac{1}{3}N^{0.25}$. Following the same perspective, in \cite{Weger} De Weger proposed another bound $d<\displaystyle\frac{N^{\frac{3}{4}}}{\vert p-q \vert}$. Later in \cite{Nitaj}, Nitaj and al. proposed the improved bound $d<\displaystyle\frac{\sqrt{6\sqrt{2}}}{6}N^{0.25}$.\\
In \cite{abcd}, Bunder and al. proposed a new improved bound and proved that $d$ can be recovered if $d<2\sqrt{2}N^{\frac{1}{4}}$.\\
More recently, in \cite{abc} Ariffin and al. proposed a new improved short decryption exponent attack for \\ $d<\displaystyle\sqrt{\frac{a^{j}+b^{i}}{2}}\left(\frac{N}{e}\right)^{\frac{1}{2}}N^{0.375}$.\\
Setting $N_{1}= N-\displaystyle\lceil\left(  \frac{a^{\frac{j}{i}}+b^{\frac{j}{i}}}{(2ab)^{\frac{j}{2i}}}+\frac{a^{\frac{1}{j}}+b^{\frac{1}{j}}}{(2ab)^{\frac{1}{2j}}}\right) \sqrt{N} \rceil+1$, they showed that $d$ can be recovered in polynomial time from the convergents of the continued fraction of $\displaystyle\frac{e}{N_{1}}$.\\
All of these attacks are bounded by the Euler totient function approximation and the size of the secret key $d$ or the size of the public exponent $e$ and are therefore limited when the public parameter $e$ is properly chosen and consequently $d$ is properly chosen.\\
Consequently, we present a generic attack based on continued fractions.
It is independent of public and private keys $e$ and $d$ respectively.\\
The attack consists on finding $\displaystyle X_{\alpha}=\left(-\alpha_{3}, \ \alpha_{3}+1 \right)$ that satisfies $\displaystyle\left\langle X_{\alpha_{3}}, \ P_{3} \right\rangle =0$ from a convergent of $\displaystyle\frac{\alpha_{i4}}{\alpha_{j4}}+\delta$. \\
When successful, it retrieves an RSA private key $d$ in $\displaystyle\mathcal{O}\left( b\log{\alpha_{j4}}\log{(\alpha_{i3}+\alpha_{j3})}\right)$.\\
For that, let's consider the following results.

\begin{Lemma} \label{lemm} \ \\ \ \rm{
$ \forall \ n$ RSA type modulus and consider the structure $\mathcal{B}_{n}(x, y)_{\mid_{x\geq 4n}}$, then $3\alpha_{j3}<\alpha_{j4}$.
}
\end{Lemma}

\begin{Proof} \ \\ \rm{
$\forall \ n=pq$, $pq<n+2$, then $q<\frac{pq}{3}<\frac{n+2}{3}$ finally $q<\displaystyle\frac{n+2}{3}$ then $3q<n+2$, and $3(q-1)<n-1$ hence $3\alpha_{j3}<\alpha_{4}$.
}
\end{Proof}
\begin{Theorem} \ \label{principal} \\ \rm{
Let $n$ be a RSA modulus and $\mathcal{B}_{n}(x, y)_{\mid_{x\geq 4n}}$ the associated hyperbola structure. There exists $\displaystyle\delta\leq\frac{1+\alpha_{j4}-\alpha_{j3}}{\alpha_{j4}\alpha_{j3}}$ and more precisely $\displaystyle 0< \frac{\alpha_{j4}-\alpha_{j3}}{\alpha_{j4}\alpha_{j3}}< \delta\leq\frac{1+\alpha_{j4}-\alpha_{j3}}{\alpha_{j4}\alpha_{j3}}$ such that $\displaystyle\frac{\alpha_{i3}}{\alpha_{j3}}$ is a convergent of \ $\displaystyle\frac{\alpha_{i4}}{\alpha_{j4}}+\delta$.
}
\end{Theorem}
\begin{Proof} \ \\ \rm{
Consider 
\begin{align*}
\displaystyle 0< \frac{\alpha_{j4}-\alpha_{j3}}{\alpha_{j4}\alpha_{j3}}< \delta\leq\frac{1+\alpha_{j4}-\alpha_{j3}}{\alpha_{j4}\alpha_{j3}}\\
\displaystyle\alpha_{j4}-\alpha_{j3}<\delta \alpha_{j4}\alpha_{j3}\leq 1+\alpha_{j4}-\alpha_{j3}\\
\displaystyle 0<\delta \alpha_{j4}\alpha_{j3}-(\alpha_{j4}-\alpha_{j3})\leq 1\\
\text{From Lemma \ref{lemm}, } 3\alpha_{j3}<\alpha_{j4}\\
\displaystyle 0<3\alpha_{j3}\left[ \delta \alpha_{j4}\alpha_{j3}-(\alpha_{j4}-\alpha_{j3})\right] \leq 3\alpha_{j3}<\alpha_{j4}\\
\displaystyle 0<2\alpha_{j3}\left[ \delta \alpha_{j4}\alpha_{j3}-(\alpha_{j4}-\alpha_{j3})\right]<\alpha_{j4}\\
\displaystyle 0<2\alpha_{j3}\left[\alpha_{j4}\alpha_{j3}-\alpha_{j4}\alpha_{j3}+ \delta \alpha_{j4}\alpha_{j3}-(\alpha_{j4}-\alpha_{j3})\right]<\alpha_{j4}\\
\displaystyle 0<2\alpha_{j3}\left[\alpha_{j3}(\alpha_{j4}+1)+ \delta \alpha_{j4}\alpha_{j3}-\alpha_{j4}(\alpha_{j3}+1)\right]<\alpha_{j4}\\
\displaystyle 0<2\alpha_{j3}\left[\alpha_{j3}\alpha_{i4}+ \delta \alpha_{j4}\alpha_{j3}-\alpha_{j4}\alpha_{i3}\right]<\alpha_{j4}\\
\displaystyle 0<(\alpha_{j3}\alpha_{i4}+ \delta \alpha_{j4}\alpha_{j3}-\alpha_{j4}\alpha_{i3})<\frac{\alpha_{j4}}{2\alpha_{j3}}=\frac{\alpha_{j4}\alpha_{j3}}{2\alpha_{j3}^{2}}\\
\text{Dividing the whole expression by } \alpha_{j4}\alpha_{j3} \text{, we have:}\\
\displaystyle0<\left|\frac{\alpha_{i4}}{\alpha_{j4}}+\delta -\frac{\alpha_{i3}}{\alpha_{j3}}\right|<\frac{1}{2\alpha_{j3}^{2}}
\end{align*}
Now applying the Legendre's theorem \cite{lang}, $\displaystyle\frac{\alpha_{i3}}{\alpha_{j3}}$ is a convergent of \ $\displaystyle\frac{\alpha_{i4}}{\alpha_{j4}}+\delta$.
 ie there exists $k\leq n$ such that $\displaystyle\left[a_{0}, a_{1}, \ldots, a_{k} \right]=\frac{\alpha_{i3}}{\alpha_{j3}}$ with $\displaystyle\frac{\alpha_{i4}}{\alpha_{j4}}+\delta=\displaystyle\left[a_{0}, a_{1}, \ldots, a_{n} \right]$.
}
\end{Proof}
\begin{Remark} \ \\ \rm{
We consider the following notation for the continued fraction of $x$, $x=\displaystyle\left[a_{0}, a_{1}, \ldots, a_{n} \right]=a_{0}+\displaystyle\frac{1}{a_{1}+\displaystyle\frac{1}{\ddots+\displaystyle\frac{1}{a_{n}}}}$. \\
The $k^{th}$ convergent of $x$ is given by $\displaystyle\left[a_{0}, a_{1}, \ldots, a_{k} \right]$ with $k\leq n$, we denote it $\displaystyle\frac{p_{k}}{q_{k}}=\displaystyle\left[a_{0}, a_{1}, \ldots, a_{k} \right]$ and the sequences $(p_{k})$ and $(q_{k})$ are computed using the recursive formulas:
\begin{align*}
p_{-2}&=q_{-1}=0, \ \ p_{k}=a_{k}p_{k-1}+p_{k-2} \ \ \text{for all } k\geq 0,\\
p_{-1}&=q_{-2}=1, \ \ q_{k}=a_{k}q_{k-1}+q_{k-2} \ \ \text{for all } k\geq 0.
\end{align*}
}
\end{Remark}

\textbf{Discussion:} \ \\
Theorem \ref{principal} gives an important result allowing us to attack any RSA modulus $n$ using the continued fraction method in a "parameters $e$ and $d$"-independent way in contrast to existing related attacks both being constrained and depending on parameters $e$ and $d$.\\
From our results, finding $\delta$ such that $\displaystyle\delta\leq\frac{1+\alpha_{j4}-\alpha_{j3}}{\alpha_{j4}\alpha_{j3}}$ is the crutial point.\\
We approach this by approximating the value of $\displaystyle\frac{1}{\alpha_{j3}}$ since by rearranging the terms, we have $\displaystyle\delta\leq\frac{1+\alpha_{j4}-\alpha_{j3}}{\alpha_{j4}\alpha_{j3}}=\frac{1+\alpha_{j4}}{\alpha_{j4}\alpha_{j3}}-\frac{1}{\alpha_{j4}}=\frac{1+\alpha_{j4}}{\alpha_{j4}}\cdot\frac{1}{\alpha_{j3}}-\frac{1}{\alpha_{j4}}>0$. \\
Knowing that $0<\displaystyle\frac{1}{\alpha_{j3}}<0.5$, 
it is not difficult to see by induction and from a simple empirical observation (see table \ref{table:alphaj3}) that the decimal part of $\displaystyle\frac{1}{\alpha_{j3}}$ has exactly $\sharp$(number of digits of $\alpha_{j3})-1$) zeros.\\
\begin{tiny}
\begin{table}[!h]
\caption{Some values of $\alpha_{j3}$ and $\frac{1}{\alpha_{j3}}$}
\begin{tabular}{c c c c c c c c c c c c}
\hline\hline
$\alpha_{j3}$ & 2 & 3 & 5 & 6 & 11 & 15 & 140 & 146 & 15756 & 1967537754 & $\cdots$ \\ 
\hline
$\displaystyle\frac{1}{\alpha_{j3}}$ & $0.5$ & $0.3$ & $0.2$ & $0.16$ & $0.09$ & $0.\underbrace{0}_{1}6$ & $0.\underbrace{00}_{2}714$ & $0.\underbrace{00}_{2}68$ & $0.\underbrace{0000}_{4}635$ & $0.\underbrace{000000000}_{9}508$ & $\cdots$ \\ 
\hline 
\end{tabular} 
\label{table:alphaj3}
\end{table}
\end{tiny}
\pagebreak \ \\
Another observation is that if $\alpha_{i4}$ or $\alpha_{j4}$ has $x$ bits then $\alpha_{i3}$ has $\frac{x}{2}$ bits. Equivalently, if $\alpha_{j4}$ has $k$ decimal digits in base $10$ then $\alpha_{j3}$ has $\displaystyle\lfloor\frac{k}{2}\rfloor+1$ decimal digits.\\
Finally $\displaystyle\frac{1}{\alpha_{j3}}$ has in its decimal part $\displaystyle\lfloor\frac{k}{2}\rfloor$ number of zeros, where $k$ is number of decimal digits of $\alpha_{j4}$. This information helps us to approximate $\displaystyle\frac{1}{\alpha_{j3}}$.\\
Therefore the following algorithm.

\subsection{Attack algorithm}
Here we present the algorithm. Its complexity analysis is given by proposition \ref{complexity} \\ \ \\
\begin{footnotesize}
\begin{algorithm}[H]\label{attack}
\KwIn{$n, \ P_{4} \in \mathcal{B}_{n}(x, y)_{\mid_{x\geq 4n}}$, \ $e$ : RSA public exponent}
\KwOut{$p$ and $q$ non trivial factors of $n$}
\textbf{Step 1}: Retrieve factors of $n$\\
\begin{algorithmic}
\STATE Initialize bound $b=10^{y}$, $y\in \lbrace 2, 3, 4, 5, 6, 7, 8, 9,... \rbrace$.
\STATE Initialize $k=\sharp \ \text{decimal digits of } n$
\STATE Initialize $\displaystyle\frac{\alpha_{i4}}{\alpha_{j4}}=\frac{x_{P_{4}}}{y_{P_{4}}}$
\WHILE{$i\leq b$}
\STATE $\delta \gets \ \displaystyle\frac{1+\alpha_{j4}}{\alpha_{j4}}\cdot\frac{i}{10^{\lfloor \frac{k}{2} \rfloor+1+\sharp \text{decimal of i}}}-\frac{1}{\alpha_{j4}}$ \ %
\STATE \IF{$\delta >0$}
 \STATE Compute convergents $\displaystyle\frac{p_{k}}{q_{k}}=\frac{\alpha_{i3}}{\alpha_{j3}}$ of $\displaystyle\frac{\alpha_{i4}}{\alpha_{j4}}+\delta$ \\
\FOR{$\displaystyle\frac{\alpha_{i3}}{\alpha_{j3}}\ \in $ convergents of  $\displaystyle\frac{\alpha_{i4}}{\alpha_{j4}}+\delta$}
\STATE  \IF{$\gcd(n, \ \alpha_{i3}+\alpha_{j3}) \ \neq \ 1, \ n$}
 \STATE \RETURN $\alpha_{i3}+\alpha_{j3}$, $\gcd(n, \ \alpha_{i3}+\alpha_{j3})$
 \ENDIF 
 \ENDFOR
 \ENDIF
\ENDWHILE
\end{algorithmic}
2 \textbf{Step 2}: Retrieve $d$ : RSA private key from $e$\\
\begin{algorithmic}
\STATE $\psi \gets \left(\alpha_{i3}+\alpha_{j3}-1\right)\left(\gcd(\alpha_{i3}+\alpha_{j3})-1\right)$
\STATE solve using extended Euclidean algorithm $ed+k\psi=1$
\RETURN $d$ 

\end{algorithmic}

\caption{\sc \textbf{Attack} Algorithm}
\end{algorithm}
\end{footnotesize}
 \ \
 \begin{Example} \ \\ \rm{
 \begin{scriptsize}
\begin{table}[h!]
\centering
\begin{tabular}{c c c c c}
\hline\hline
RSA modulus & Size (bits) & $\delta$ & $\frac{\alpha_{i3}^{'}}{\alpha_{j3}^{'}}$ & $\gcd(\alpha_{i3}^{'}+\alpha_{j3}^{'}, \ n)$ \\ [0.5ex]
\hline
$14893$ & $17$ & $\frac{7}{10^{3}}$ & $\frac{141}{140}$ & $281$ \\
$439007603$ & $30$ & $\frac{969}{5\cdot 10^{7}}$ & $\frac{122935301}{122932918}$ & $13931$ \\ [1ex]
\hline
\end{tabular}
\label{table:experiments}
\end{table}
\end{scriptsize}
 }
 \end{Example}
\newpage
\begin{Remark} \ \\ \rm{
This attack can be parallelized in the following way:
In the while loop, while iterating on different values of $i$ up to $b$, discretize the interval in different steps.\\
That said, assume for example we want to distribute the work on 2 CPUs, then:\\
in CPU$_{1}$ go from $0$ to $b$ with step 2: $0, 2, 4, \cdots,$ up to $b$.\\
in CPU$_{2}$ go from $1$ to $b$ with step 2: $1, 3, 5, \cdots,$ up to $b-1$. 
}
\end{Remark}
\begin{Proposition}\label{complexity} \  \\ \rm{

This algorithm \ref{attack} has a complexity in $\displaystyle\mathcal{O}\left( b\log{\alpha_{j4}}\log{(\alpha_{i3}+\alpha_{j3})}\right)$.
}
\end{Proposition}
\begin{Proof} \ \\ \rm{
step 1, initializations are done in constant time $\mathcal{O}(1)$, $\displaystyle\frac{\alpha_{i4}}{\alpha_{j4}}+\delta$  has $\mathcal{O}(\log{}\alpha_{j4})$ convergents and the computation of the greatest common divisor of $n$ and $\alpha_{i3}+\alpha_{j3}$ has complexity in $\mathcal{O}(\alpha_{i3}+\alpha_{j3})$ since   $\min\left(n, \ \alpha_{i3}+\alpha_{j3} \right)=\alpha_{i3}+\alpha_{j3}$, thus a total of $\mathcal{O}\left(b\log{\alpha_{j4}}\log{(\alpha_{i3}+\alpha_{j3})} \right)$. \\
Step 2 has complexity in $\mathcal{O}\left( \log{}e\right)$ since $1<e<\psi$ implies $\min(e, \ \psi)=e$.\\
Now $\max(b\log{\alpha_{j4}}\log{(\alpha_{i3}+\alpha_{j3})}, \ \log{}e)=b\log{\alpha_{j4}}\log{(\alpha_{i3}+\alpha_{j3})}$, hence \\ $\mathcal{O}\left(\text{step 1}+\text{step 2}\right)=\displaystyle\mathcal{O}\left( b\log{\alpha_{j4}}\log{(\alpha_{i3}+\alpha_{j3})}\right)$.
}
\end{Proof}

\section*{\textbf{Conclusion}}  \rm{
We have presented new arithmetical and algebraic results following the work of Babindamana and al. in \cite{Gilda} on hyperbolas and describe an approach to attacking a RSA-type modulus based on continued fractions in the new results, different from Wiener's attack approach which depends on the parameters $e$ and $d$. When successful, this attack has a complexity in 
$\displaystyle\mathcal{O}\left( b\log{\alpha_{j4}}\log{(\alpha_{i3}+\alpha_{j3})}\right)$.\\
The particularity of this attack is that it is independent and not bounded by the size of the private key $d$ nor the public exponent $e$ compared to Weiner's attack  and its improvements. 
As future work, this method should be thoroughly investigated to find the best bounds for $\delta$, cases of modulus with faster $\frac{\alpha_{i3}^{'}}{\alpha_{j3}^{'}}$ retrivial and reduce $b$ for a polynomial time overall complexity. 
}


\begin{thebibliography}{01}
\bibitem{Gilda} Regis F. Babindamana, Gilda R. Bansimba, Basile G. R. Bossoto (2022). \emph{"Lattice Points on the Fermat Factorization Method"}, Journal of Mathematics, vol. 2022, Article ID 6360264. 

\bibitem{Rech} Gilda R. Bansimba, Regis F. Babindamana, Basile G. R. Bossoto (2021). \emph{Some Arithmetical properties on Hyperbola}, JP Journal of Algebra, Number Theory and Applications, Vol. 50, no. 1, P. 45-100.

\bibitem{abc} Abubakar, S. I., Ariffin, M. R. K. and Asbullah, M. A. (2019). \emph{A New Improved Bound for Short Decryption Exponent on RSA Modulus N = pq using Wiener’s Method}, Malaysian Journal of Mathematical Sciences 13(S) April: 89–99 

\bibitem{abcd} Bunder, M. and Tonien, J. (2017). \emph{A New Attack on the RSA Cryptosystem Based on Continued Fractions}, Malaysian Journal of Mathematical Sciences 11(S) August: 45 - 57 (2017).

\bibitem{Wiener} Wiener, M. J. (1990). \emph{Cryptanalysis of short rsa secret exponents}, IEEE Transactions on Information theory, 36(3):553–558.

\bibitem{Weger} De Weger, B. (2002). \emph{Cryptanalysis of rsa with small prime difference}. Applicable Algebra in Engineering, Communication and Computing, 13(1):17–28.

\bibitem{Nitaj} Nitaj, A. (2013). \emph{Diophantine and lattice cryptanalysis of the rsa cryptosystem}. In Artificial Intelligence, Evolutionary Computing and Metaheuristics, pages
139–168. Springer.

\bibitem{coppersmith} Don Coppersmith (1997), \emph{Small Solutions to Polynomial Equations, and Low Exponent RSA Vulnerabilities}. Journal of Cryptology,  10(4), 233–260.

\bibitem{lang} S. Lang (1966), \emph{Introduction to Diophantine Approximations}, Addison-Wesley Pub. Co.




\end{thebibliography}
\end{document}